# The Geometry of Spaceborne Synthetic Aperture Radar

P. H. Robert Orth

**Abstract**

The paper defines the new coordinate system that was developed in 1977-78 for the world's first digital synthetic aperture radar (SAR) ground processor for images from the Seasat-A satellite. The coordinate system is for the range-Doppler paradigm in the context of a spaceborne platform orbiting a rotating planet. The mathematical expressions for the azimuth FM rate, isodoppler lines, target illumination trajectories and antenna attitude determination from Doppler centroid measurements are derived. The method for transforming the SAR images from that SAR digital signal processor that used these parametric inputs is also presented. The paper concludes with a report of the measurement of the map accuracy of the resulting images.

**Introduction**

It is approaching 40 years since the first digitally processed spaceborne Synthetic Aperture Radar (SAR) image was produced [1] in November 1978, from the NASA Seasat-A [2] (or Seasat) data. In September 2014 that achievement was awarded the I.E.E.E Milestone in Electrical Engineering and Computing [3]. The signal processing algorithms developed for the processor and their subsequent development have been extensively described elsewhere [4].

Until 1977 there was no civilian spaceborne SAR geometry model available for that first digital processor to provide the Doppler shift data. This paper describes the original geometric model devised at that time for spaceborne SAR, providing key parameters for signal processing and the image-to-ground map transformations. Work on developing the model was started and continued in a series of workshops under the auspices of Communication Research Canada and the Canada Centre for Remote Sensing. At that time the decision was made to build a processor that would compress azimuth data to zero Doppler.

Therefore, following that decision, work began on the definition of the coordinate systems required, the transformation equations linking them, and completed with the expressions critical for the SAR signal processing parameters. The final product of the SAR processor was defined to be an image on a standard map projection (e.g. UTM) on the WGS 84 geoid, in order to measure the spatial accuracy of a selected target relative to it's published map location. Emphasis was also placed on the usefulness of SAR images on other conventional map projections, to tie in with remote sensing images such as Landsat.

In what follows specific results for Seasat have not been included except where specifically mentioned. For clarification vectors have a single underscore, and matrices a double underscore. Vector dot products are denoted $\underline{v}\,\underline{v}^T = \underline{v}^T\underline{v}$. In keeping with the fact the slant range is a scalar, the term Doppler speed is used and not Doppler velocity.

**Outline of Approach**

The geometrical aspects of spaceborne SAR processing can be divided into three main categories.
(a)     The determination of the slant range or radial distance to any point target fixed to the surface of the earth, as measured from the satellite (ideally, from the phase centre of the radar antenna).
(b)     The determination of the placement of the radar beam in the reference coordinate system fixed to the satellite. Only when the point target referred to in (a) has the appropriate coordinates will the radar beam illuminate the point target, and so contribute to the measured signal. In general, because of the variation of spacecraft attitude angles together with the effect of earth rotation, the target may not be illuminated at its point of closest approach (minimum slant range, or point of zero doppler) to the spacecraft.
(c)     The determination of sufficiently precise satellite orbit data to provide the parameters for (a) and (b) above. In this sense a definite separation exists in the geometrical problems encountered, in that the first two aspects (a) and (b) form the geometric elements of the SAR processor, and proceed independently from the formulation of the best orbit solution. Included in this last aspect is the appropriate shape of the earth and local elevations relative to the earth ellipsoid. The bulk of the effort involved the formulation of solutions to (a) and (b), based on the availability of orbit data provided with the Seasat data. This procedure is analogous to the operational mode followed in any spaceborne remote sensing, where separate measurements of spacecraft orbital position (e.g. GPS) are supplied for image processing and annotation.

**Preliminary Geometrical Concepts**

The along-track direction of a section of data recorded from the SAR corresponds precisely to an orbital time axis. The returns from each successive radar pulse (at frequency PRF or Pulse Repetition Frequency) are recorded in "slow" time, line-by-line in a "*signal memory*". Each line is divided into slant range cells according to the "fast" time of digitization. It is assumed that range compression has been applied to this range data.
It is possible to construct a second memory namely "*image memory*", with exactly the same slow time and fast time coordinates as signal memory. An observer fixed to the spacecraft (in the "*satellite coordinate system*") considers the passage relative to a fixed point on the ground (in the *earth-centred earth-fixed or ECEF coordinate system*) [5]. The motion of the point relative to the observer is the vector sum of motions due to the spacecraft velocity in the orbital plane, and the velocity due to earth rotation. This resultant radial motion is seen to approach the spacecraft, pass through a minimum and then recede. The orbital time corresponding to the minimum slant range defines a unique slow time in the second memory, while the minimum slant range defines a unique fast time. Since each ground point has such a unique location, the second memory is effectively an image memory. Furthermore, since the time coordinate axes are identical to signal memory it is possible to locate the target's time of zero slant range motion, or zero Doppler frequency, in the signal memory.

Of particular interest is the set of coordinate values in image memory corresponding to minimum recorded fast time (slant range), for various values of slow time. To the observer on board the spacecraft, these coordinates correspond to points on the earth such that all lines of constant fast time in image memory are parallel to the sub-satellite track. Equally, a set of targets aligned parallel to the sub-satellite track will be located along lines of constant fast time, according to that slant range at the time of zero Doppler.

Consider now a radar beam (in the *radar coordinate system*) that is oriented with its widest extent perpendicular to the sub-satellite track (which is the earth-surface intersection of a line joining the satellite to the earth centre), over a particular section of the orbit. In this case the SAR measurements in signal memory, at some particular slow-time $t_k$, contain contributions from targets that are at zero Doppler, passing through their minimum slant ranges. The corresponding image memory slow-time coordinates for all these target points perpendicular to the sub-satellite track is defined as $t_k$.

Next consider a beam that is slightly yawed forward so that these same targets are illuminated at a time previous to $t_k$. In signal memory the result is that these targets contribute to data cells at a time previous to $t_k$ and, taking into account the concomitant change in slant range, also contribute to various fast-time cells, depending on the width of the beam.

However, in image memory, whose coordinates are the fast-time and slow-time at zero Doppler, the same coordinates as in the non-yawed case represent the target.

In the absence of spacecraft attitude variation the SAR beam is not necessarily perpendicular to the sub-satellite track, since the radar antenna is fixed to the spacecraft whose velocity vector deviates from the sub-satellite track direction, except when the orbit parallels the earth rotation velocity. This displacement of the beam leads again to the illumination of the targets being different from zero Doppler. Throughout the orbit, this beam displacement varies slowly, and must be combined with the effects of yaw, pitch and roll to produce a final illumination direction for the beam.

The trajectory of a point target in signal memory and the time of its illumination are readily calculated in the satellite coordinate system. Since the slow-time axis of image memory parallels the sub-satellite track, one axis of this satellite coordinate system is chosen to be in the sub-satellite track direction. The conventional choice for the $z$ axis is alignment with the line joining spacecraft centre of mass to the earth's centre of mass. The third axis is chosen to form a rectangular coordinate system. Remembering that lines of latitude and longitude intersect at right angles, this satellite coordinate system has the $xy$ axes rotated in some specific manner about the $z$ axis with respect to the lines of latitude and longitude at the sub-satellite point. A line of points perpendicular to the real track direction are seen, by an observer in the satellite coordinate system, to cross the $x$ axis (at $y = 0$) at precisely the same instant. Motion of the points is from positive to negative along the $y$ axis, and importantly have zero velocity in the $y$ axis direction. This fact is used in what follows to determine the orientation of the $xy$ axis.

The satellite coordinate system trajectory of a target having specific zero Doppler (image memory) coordinates is calculated using the method of intersecting surfaces [7]. A unique point in the satellite coordinate system is identified by finding the intersection point of the surface of constant Doppler, the radar spherical wave with radius equal to slant range, and the representation of the earth's surface. Considerable simplification of the resulting expressions results from aligning the $y$ axis to the real track direction rather than to the satellite velocity vector. This simplification arises primarily in the transformation of image memory coordinates to the satellite coordinate system. Thus, at zero Doppler, for the case of negligible rate of change $dH/dt$ ($\dot{H}$) of orbit radius $H$ (see **Transformation Equations** below), the $y$ component of the target is always zero. For any other orientation of the $xy$ axes, this would not be the case. It will be seen later that when $\dot{H}$ is not negligible, the main effect is displacement of the zero Doppler position from $y = 0$ by an amount proportional to $\dot{H}$. This does not affect the time of zero Doppler, that is, the image memory coordinates.

A number of specific features of the approach taken are as follows:

\*       The formulation of SAR geometry in terms of measured spacecraft ECEF coordinates (e.g., latitude, longitude and orbital altitude as a function of time).

\*       With the slow-time axis lined up with real track direction, the zero doppler line is always perpendicular to slow-time axis, and provided compression is to zero doppler, look registration is automatically achieved.

\*       Separation of the earth shape from orbit trajectory allows the straightforward inclusion of effects of the non-spherical earth.

\*       The position of the beam illumination pattern in signal memory is clearly dependent on combined effects of earth rotation and satellite altitude.

\*       Rectangles on the ground oriented at right angles to the sub-satellite track map into rectangles in image memory. Ground lines not parallel or perpendicular to the sub-satellite track are, however, curved in image memory due to the effect of slant range.

\*       Ready formulation of measurement of attitude parameters from Doppler centroid data.

## Coordinate Systems

The coordinate systems used for this work are defined in this section.

### Signal Memory

The signal memory coordinate system is defined as a right-handed Cartesian system with fast time $\tau$ ($\tau = 0$ is the fast time at which the radar pulse is transmitted) along the $+x$ axis; slow time $\pm\eta$ (with $\eta = 0$ at orbit time $t_k$) along the $\pm y$ axis; and SAR signal intensity S, after range compression, along the $+z$ axis. The coordinate system is shown in Figure 1.

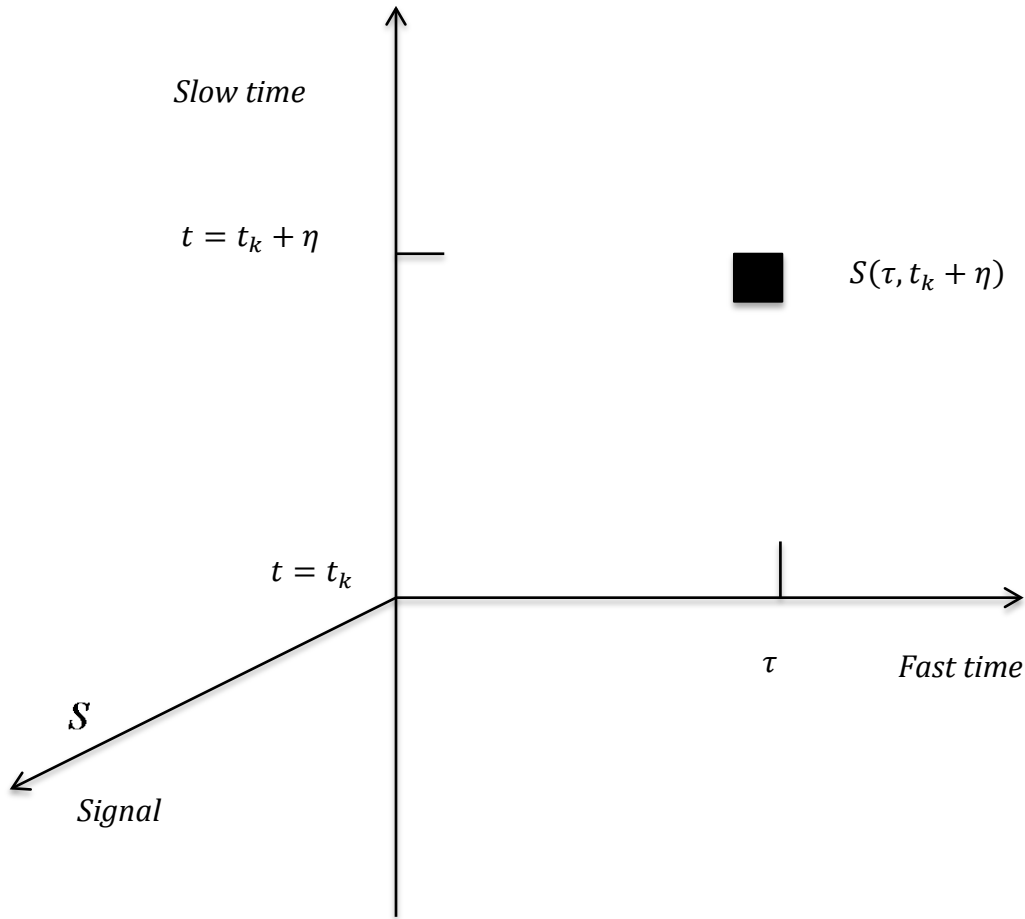

Figure 1.     **Signal memory.**

The fast time axis is divided into slant range gates or bins. Cell widths in the slow time direction are dependent on the PRF. For the nominal Seasat sub-satellite track speed of 6.66 km/sec, 100 km along the track corresponds to $\Delta\eta = 15$ secs, while 25m along track corresponds to $\Delta\eta = 3.75$ msecs.

**Image Memory**

The image memory coordinate system is defined as a right-handed Cartesian system with the fast time $\tau$ ($\tau = 0$ being the time of radar pulse transmission) of zero doppler frequency (or closest approach to the spacecraft for any point target fixed on the earth) placed along the $+x$ axis; slow time $\pm\eta$ of zero Doppler frequency along the $\pm y$ axis (again referenced at $\eta = 0$ to an orbit time $t_k$). The reflectivity $\mathcal{I}$ of a target having zero Doppler at $(\tau, \eta)$ is plotted along the $z$ axis. The coordinate system is shown in Figure 2. The analysis of geometric effects must supply the following information to the SAR processor, in order that contents of signal memory can be processed to obtain the target reflectivity $\mathcal{I}$ at any $(\tau, t_k)$ in image memory:

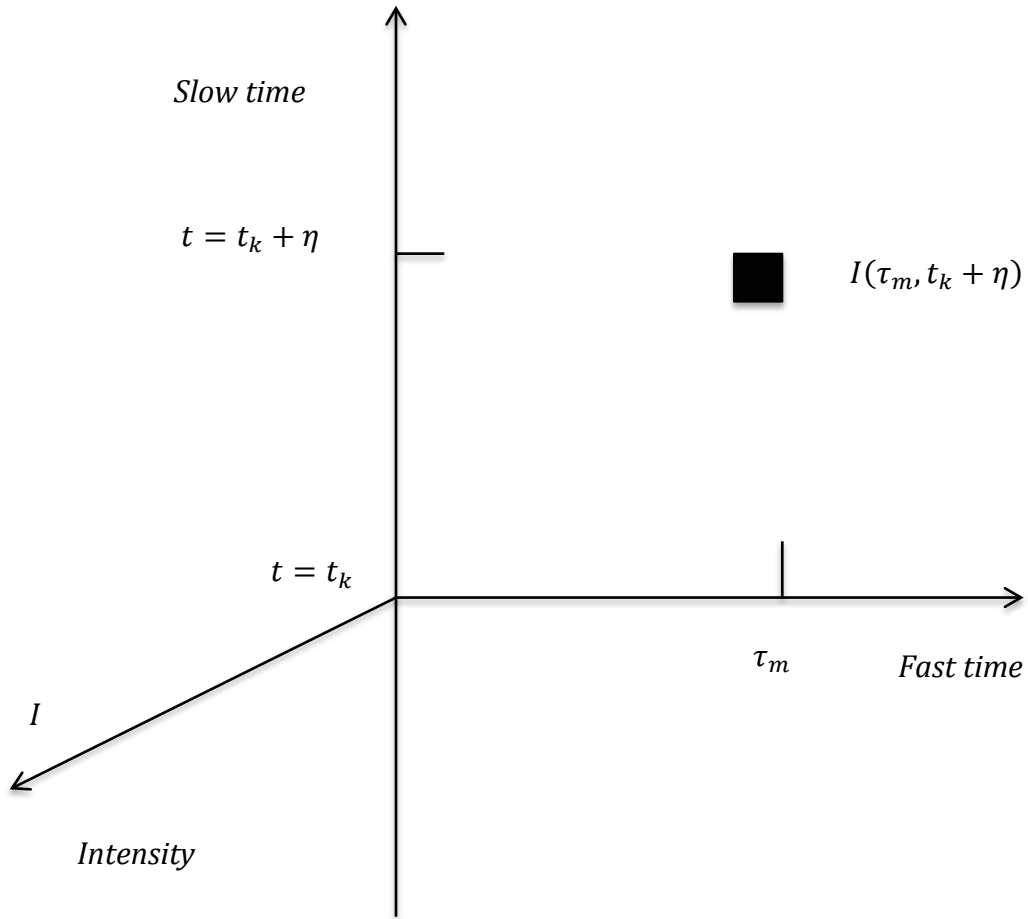

Figure 2.   **Image memory.**

The contents of image memory are the result of azimuth compression of signal memory data to zero Doppler frequency. As in the case of signal memory, the slow time and fast time axes have cell sizes associated with them. For Seasat these corresponded to ground range and azimuth (along track) resolution of 25m, and a slant range cell width of 8.4 m or 56 nsec.

(a) The trajectory in signal memory of a point located at fast time $\tau_m$, slow time (orbit time) $t_k$ in image memory;
(b) The Doppler frequency of the target along its signal memory trajectory;
(c) The extent to which the SAR antenna illuminates this trajectory, taking into account measured spacecraft roll, pitch and yaw, and the effect of earth rotation.
(d) An image memory resampling procedure to correct for slant range and project onto a user specified map coordinate system, at some final scale (e.g. 1:250,000).

**Inertial Coordinate System**

This non-rotating coordinate system is required to represent the motion of the satellite with respect to the planet Earth [5][6]. The origin is placed at the centre of the Earth, and the principal plane of reference is an extension of the Earth's equatorial plane. The trajectory of the satellite is the "nominal track" [6]. It deviates from the "ground track" due to the Earth's rotation. In order to calculate this deviation angle $H_e$ for Seasat, the method used was based on earlier work on LANDSAT 1 (ERTS)[6].
The Seasat orbit had a nominal eccentricity $e$ of 0.0008 and a maximum of 0.002 and the difference between the major and minor semi-axes is approximately 2 metres. The foci are separated by 11.476 km with the Earth Coordinate system centre at the prime focus.

**Satellite Coordinate System**

The SAR measurements are made with respect to a coordinate system centred at the satellite. In particular, target slant range refers to the distance between the antenna phase centre and a target fixed to the earth, but moving relative to the satellite. Targets on the earth (assuming a radius $r_e$ at the sub-satellite point) that pass through the sub-satellite point $(0,0, H - r_e)$ do so parallel to the $y_s$ axis, and proceed from $+y_s$ to $-y_s$. The $x_s$ axis is chosen to form a right-handed coordinate system. This coordinate system is shown in Figure 3. In this system, the SAR antenna, with a clock angle of 90°, points towards the $-x_s$ axis (to the right or starboard) of the real track ($y_s$ axis) direction [7]. A clock angle of 270° points the radar in the $+x_s$ axis, to the left (port) of the real track direction. The Seasat antenna, neglecting spacecraft roll, pitch and yaw, had an inclination of 20° from the $z_s$ axis. The real track direction can always be plotted on a Cartesian coordinate system centred at the sub-satellite point, having axes parallel to the longitude and latitude lines. For this reason, the $y_s$ and $x_s$ axes are obtained by rotating these latitude, longitude lines so that one of them aligns with the real track direction. Furthermore, the sub-satellite point velocity can be immediately obtained from the rates of change of orbit longitude and latitude.
This exemplifies the connection between the satellite coordinate system and the earth coordinate system. The satellite coordinate system is an intermediate coordinate system, facilitating the development of expressions for slant range, Doppler frequency, and the transformation from image memory to the earth coordinate system.

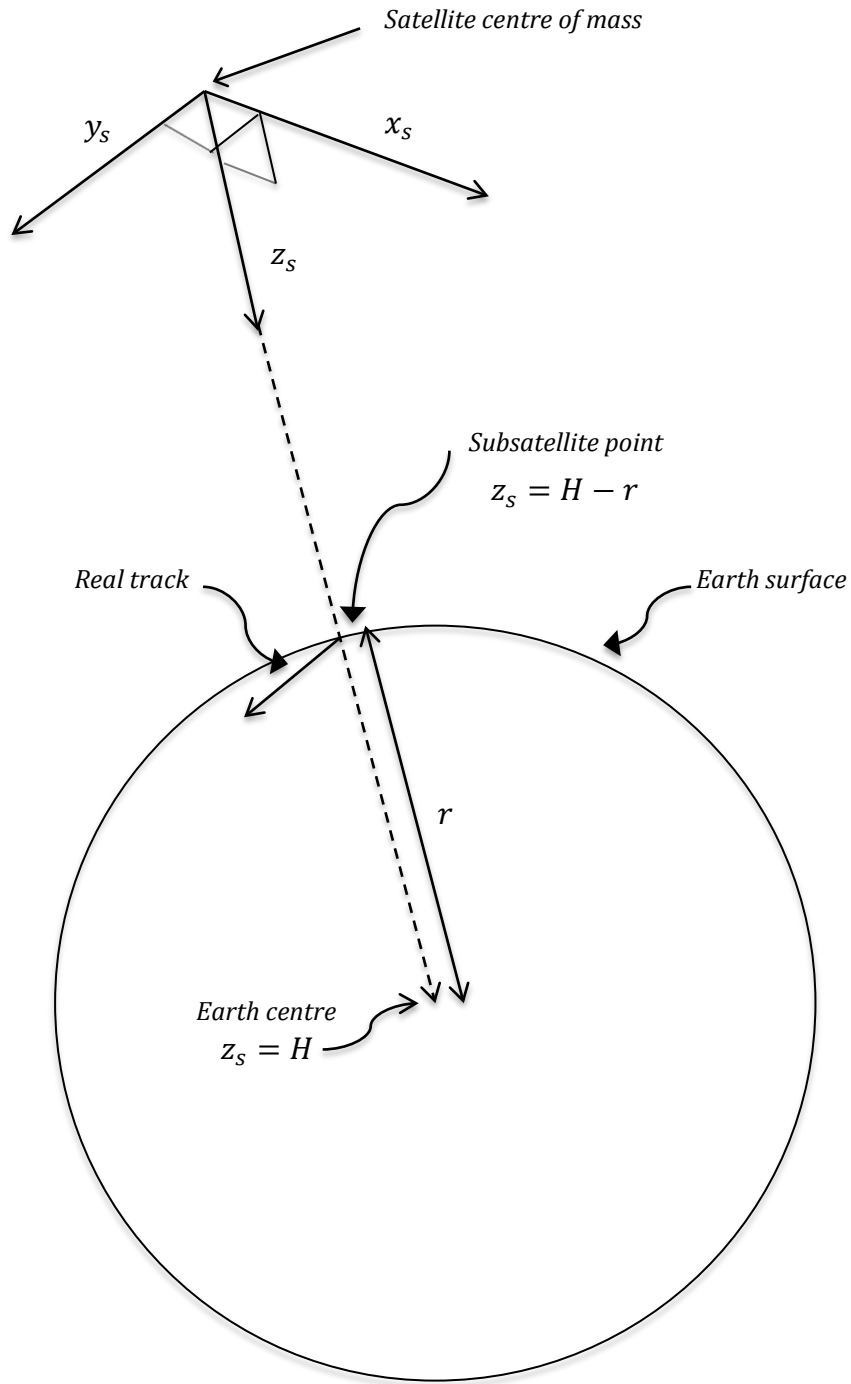

Figure 3. **Satellite Coordinate System.**

The satellite coordinate system is defined as a right-handed Cartesian system $(x_s, y_s, z_s)$ with the origin at the centre of mass of the satellite. The $z_s$ axis is directed along the line joining the satellite centre of mass to the earth centre of mass located at $z_s = H$. The $x_s y_s$ plane is normal to the $z_s$ axis, with the $y_s$ axis directed along the satellite real track.

## Earth-Centred Earth-Fixed (ECEF) Coordinate System

The ECEF coordinate system is fixed to the rotating earth [5] (Figure 4). A radial distance $r_e$ of the earth's surface at angular coordinates $\theta_e, \phi_e$ describes the shape of the earth. Here $\theta_e, \phi_e$ represent the azimuthal and polar angle of a vector in the earth coordinate system $\begin{bmatrix} x_e \\ y_e \\ z_e \end{bmatrix} = r_e(\theta_e, \phi_e) \begin{bmatrix} \cos\theta_e \cos\phi_e \\ \sin\theta_e \cos\phi_e \\ \sin\phi_e \end{bmatrix}$ and are related to latitude and longitude by the following rules:

If $\theta_e < 0$, then coordinate is at $|\theta_e|$ West Longitude
If $\theta_e > 0$, then coordinate is at $|\theta_e|$ East Longitude
If $90°-\phi_e > 0$, then coordinate is at $|90°-\phi_e|$ North Latitude
If $90°-\phi_e < 0$, then coordinate is at $|90°-\phi_e|$ South Latitude

In this coordinate system the satellite position vector has angular coordinates $\hat{\theta}, \hat{\phi}$ at a radial distance $H$, which depends in general on $\hat{\theta}, \hat{\phi}$ due to orbit eccentricity. In what follows the coordinates of the satellite are denoted by the Cartesian triplet $(H\cos\hat{\theta}\sin\hat{\phi}, H\sin\hat{\theta}\sin\hat{\phi}, H\cos\hat{\phi})$. Note that the sub-satellite point has a radial distance $r_e(\hat{\theta}, \hat{\phi})$ in the earth coordinate system.

For Seasat the values of $H, \hat{\theta}$, and $\hat{\phi}$ were obtained from the Definitive Orbit Record (DOR) and their time derivatives from a polynomial fit to the DOR data, if they were not found on the DOR. The DOR formed part of the Seasat data tape. For simulations an orbit model was used [5] [6]. These parameters are required to determine the elements of the transformation necessary to map image memory to the earth coordinate system, to determine the target trajectories in signal memory and the azimuth compression parameter, and to determine which section of this trajectory is illuminated by the radar beam.

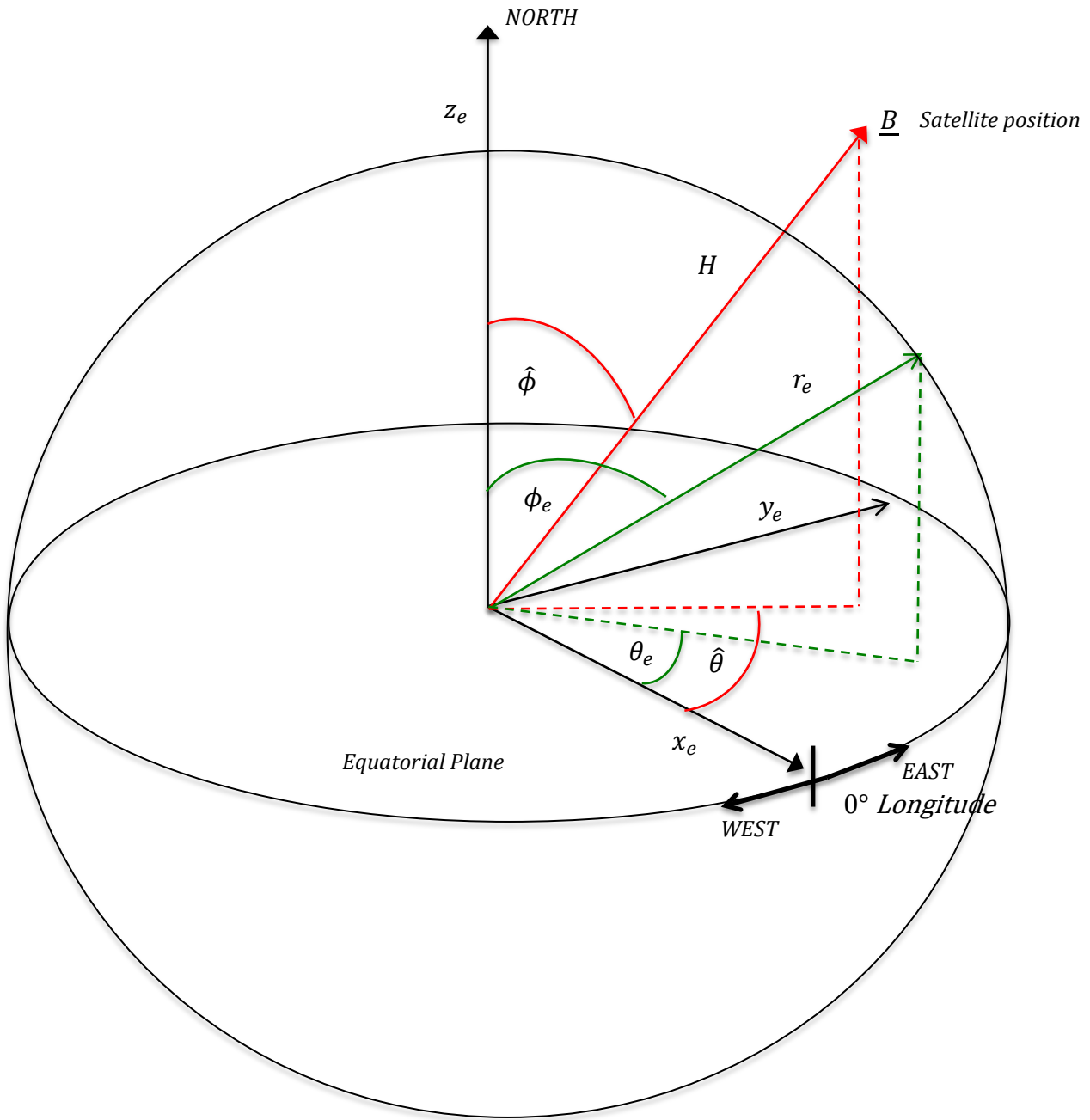

Figure 4.  **Earth-Centred Earth-Fixed (ECEF) Coordinate System.**

The ECEF is described by a right-handed coordinate system $(x_e, y_e, z_e)$ centred at the centre of mass of the earth. The $z_e$ axis is the north polar axis of the earth, while the $x_e$ axis in the equatorial plane intersects the earth's surface at 0° longitude. The $y_e$ axis in the equatorial plane is at right angles to $x_e$ to form a right-handed coordinate system.

**Radar Coordinate System**

The SAR antenna is located in a right-handed Cartesian coordinate system $(x_a, y_a, z_a)$ with origin at the centre of mass of the satellite (Figure 5). This approximation neglects the vector distance of a suitably defined antenna phase centre from the spacecraft centre of mass, and was sufficiently accurate for Seasat. The "nominal" velocity vector [6] is the velocity in the inertial coordinate system, in which the spacecraft moves with the earth seen rotating from west to east. This distinguishes it from the ECEF system that is fixed to the earth. It is in the inertial coordinate system where the satellite moves in a plane as it orbits in space. The antenna illumination pattern is on a unit sphere in this coordinate system, and points on this unit sphere are determined by $\alpha$ the "across track" angle, and $\beta$, the "along track" angle. These angles are related to the Cartesian coordinates $(x_a, y_a, z_a)$ by $\tan \alpha = y_a/z_a$ and $\tan \beta = x_a/z_a$. This coordinate system is defined to be independent of spacecraft attitude angles. The -3dB, 0dB loci in the Seasat SAR beam had $\beta = \pm 0.516°$ and $0°$ respectively [2]. Any vector along these loci illuminates a certain point on the earth's surface, and hence signal memory, depending on the spacecraft attitude angles. For Seasat the Definitive Attitude Record (DAR) contained the roll, pitch and yaw angle measurement data. The attitude angles are defined in a right-handed coordinate system where the roll axis is aligned to the spacecraft nominal velocity vector, the yaw axis is along the line joining the spacecraft centre of mass to the earth centre of mass, and the pitch axis starboard of the roll axis [8]. Positive angular displacement directions for this coordinate system are shown in Figure 6. The orientation of the coordinate system is derived from the satellite coordinate system $(x_s, y_s, z_s)$ by a rotation about the $z_s$ axis to align the satellite coordinate system velocity vector to the nominal track velocity, followed by a rotation of $90°$ to align the resulting $y$ axis to the pitch axis. Following a further set of rotations to account for sequential roll, pitch and yaw, two final rotations about the resulting $z$ and $x$ axes (in that order) take into account clock and elevation angle and lead to the radar coordinate system.

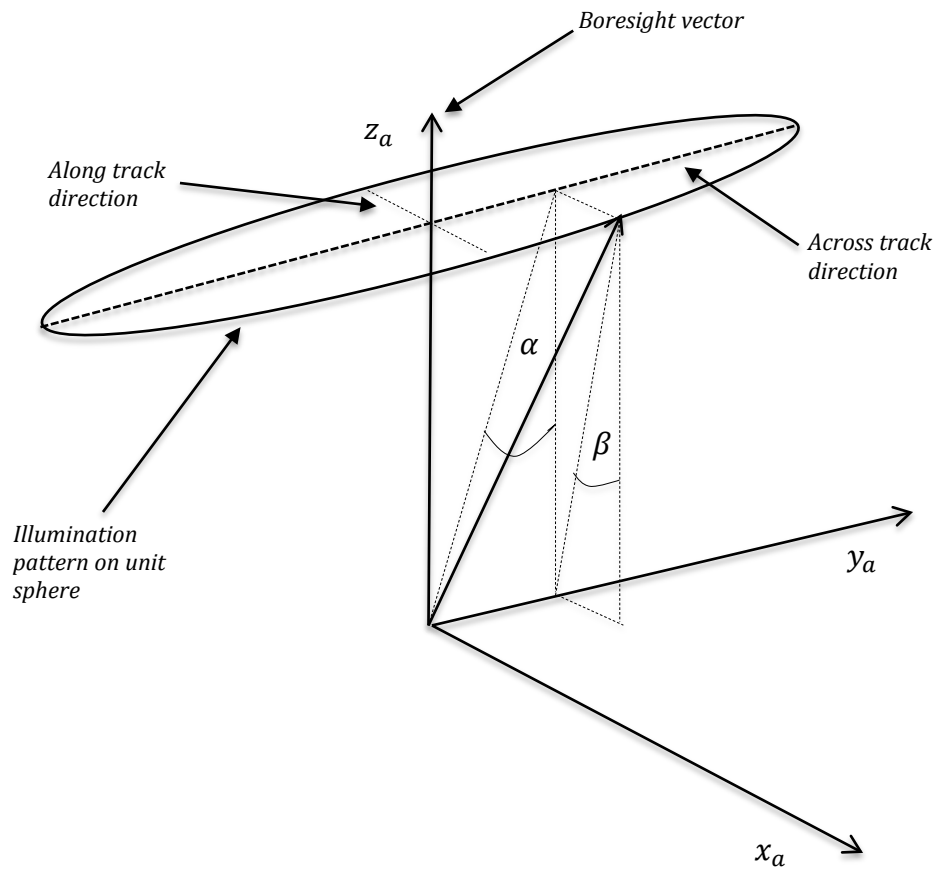

Figure 5. **Radar Coordinate System.**

The $z_a$ axis points along the boresight vector, and the $x_a$ axis parallel to the "along track" direction. This direction is intrinsic to the antenna, and is the direction of the spacecraft velocity for an antenna pointing toward either port or starboard. The spacecraft velocity is in the inertial coordinate system. The SAR beam is narrowest in the along track direction. The $y_a$ axis is in the "across track" direction.

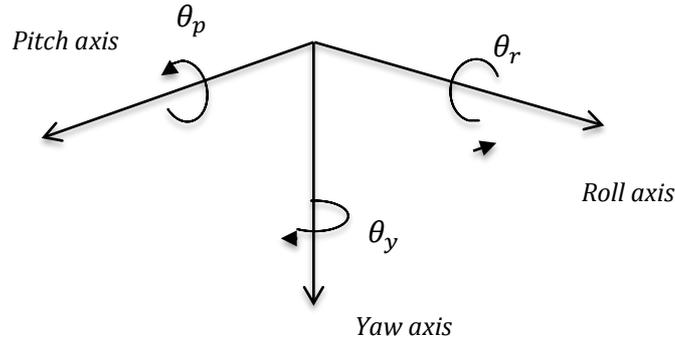

Figure 6

**Yaw, Pitch and Roll Axes.**

Positive angular displacement directions for the yaw, pitch and roll axes [8]

**Transformation Equations**

The coordinate systems of defined above form the essential framework within which key spaceborne SAR variables are expressed. The way in which these variables transform from one coordinate system to another forms the substance of this section. For example, the $(\tau, \eta)$ coordinates of a point in image memory must be related to its latitude and longitude on the earth. Furthermore, to determine the intensity of the SAR beam that illuminates a certain portion of a trajectory in signal memory requires the transformation of SAR beam vectors into the satellite coordinate system.

**Earth Coordinate System to Satellite Coordinate System**

The transformation of a vector $\underline{X}_e$ in the earth coordinate system into the vector $\underline{X}_s$ in the satellite coordinate system is accomplished by first translating to the satellite position vector $\underline{B}$ (Figure 4). This is followed by rotating the coordinate system using a rotation matrix [8] $\underline{\underline{M}}$ to align the $y_s$ axis to the real track direction, and the $z_s$ axis to point towards the earth centre along $-\underline{B}$. The result is that

$$\underline{X}_s = \underline{\underline{M}}(\underline{X}_e - \underline{B}) \qquad (1)$$

where

$$\underline{B} = H \begin{bmatrix} \cos\hat{\theta}\sin\hat{\phi} \\ \sin\hat{\theta}\sin\hat{\phi} \\ \cos\hat{\phi} \end{bmatrix} \quad (2)$$

and

$$\underline{\underline{M}} = \begin{bmatrix} \cos\hat{\phi}\cos\hat{\theta}\cos\psi + \sin\hat{\theta}\sin\psi & \cos\hat{\phi}\sin\hat{\theta}\cos\psi - \cos\hat{\theta}\sin\psi & -\sin\hat{\phi}\cos\psi \\ -\cos\hat{\phi}\cos\hat{\theta}\sin\psi + \sin\hat{\theta}\cos\psi & -\cos\hat{\phi}\sin\hat{\theta}\sin\psi - \cos\hat{\theta}\cos\psi & \sin\hat{\phi}\sin\psi \\ -\sin\hat{\phi}\cos\hat{\theta} & -\sin\hat{\phi}\sin\hat{\theta} & -\cos\hat{\phi} \end{bmatrix} \quad (3)$$

The angle $\psi$ is found by requiring that the $y_s$ axis be aligned to the real track direction. Any point $\underline{X}_e$ fixed to the earth is seen to have a varying position $\underline{X}_s$ in the satellite coordinate system since both the satellite position vector $\underline{B}$ and the rotation matrix $\underline{\underline{M}}$ depend on orbital time $t$.

The reverse transformation is

$$\underline{X}_e = \underline{\underline{M}}^T \underline{X}_s + \underline{B} \quad (4)$$

The angle $\psi$ is determined as follows. The slant range squared to any point $\underline{X}_s$ is given by $\underline{X}_s^T \underline{X}_s = x_s^2 + y_s^2 + z_s^2 = (c\tau/2)^2$ with $\tau$ being the (fast) time taken for the radar beam to traverse the round trip distance $\sqrt{\underline{X}_s^T \underline{X}_s}$. The Doppler speed of any point $\underline{X}_s$ is given by

$$d\left(\sqrt{\underline{X}_s^T \underline{X}_s}\right)/dt = \left(\underline{\dot{X}}_s^T \underline{X}_s\right)/\left(\sqrt{\underline{X}_s^T \underline{X}_s}\right). \quad (5)$$

In components Equation (5) is

$$v_d = \left(\underline{\underline{M}}\,\underline{\dot{B}}\right)^T \underline{X}_s = -(m_x x_s + m_y y_s + m_z z_s)/(c\tau/2) \quad (6)$$

where

$$\begin{bmatrix} m_x \\ m_y \\ m_z \end{bmatrix} = H \begin{bmatrix} -\omega_2 \\ \omega_1 \\ \dot{H}/H \end{bmatrix} \quad (7)$$

are the three components of $-\underline{\underline{M}}\,\underline{\dot{B}}$; and

$$\begin{bmatrix} \omega_1 \\ \omega_2 \\ \omega_3 \end{bmatrix} = \begin{bmatrix} \dot{\hat{\phi}}\sin\psi + \dot{\hat{\theta}}\sin\hat{\phi}\cos\psi \\ \dot{\hat{\phi}}\cos\psi - \dot{\hat{\theta}}\sin\hat{\phi}\sin\psi \\ -\dot{\psi} + \dot{\hat{\theta}}\cos\hat{\phi} \end{bmatrix} \quad (8)$$

Here $\underline{\underline{\dot{M}}}\,\underline{\underline{M}}^T$ is a skew symmetric matrix [8] so that $\left(\underline{\underline{\dot{M}}}\,\underline{\underline{M}}^T \underline{X}_s\right)^T \underline{X}_s = 0$.

The rate of change of $\underline{X}_s$ with respect to time is given by $\underline{\dot{X}}_s = \underline{\underline{\dot{M}}}\,\underline{\underline{M}}^T \underline{X}_s - \underline{\underline{M}}\,\underline{\dot{B}}$ for points fixed to the earth $(\underline{\dot{X}}_e = 0)$. This describes a velocity field where the velocities are dependent on position $\underline{X}_s$. In component form

$$\underline{\dot{X}}_s = \begin{bmatrix} -\omega_3 y_s - \omega_2(H-z_s) \\ \omega_3 x_s + \omega_1(H-z_s) \\ \omega_1 y_s + \dot{H} - \omega_2 x_s \end{bmatrix} = \begin{bmatrix} \dot{x}_s \\ \dot{y}_s \\ \dot{z}_s \end{bmatrix} \quad (9)$$

For the real track direction to be aligned with the $y_s$ axis, those points fixed to the earth must cross the $x_s$ axis (i.e. at $y_s = 0$), with $\dot{x}_s = 0$ which in turn requires that $\omega_2 = 0$ in Equation (8), and this results in

$$\tan\psi = -\dot{\hat{\phi}} / \left(-\dot{\hat{\theta}}\sin\hat{\phi}\right) \quad (10)$$

This choice for $\psi$ therefore ensures that the real track direction is aligned with the $y_s$ axis, with all earth-fixed points crossing the $x_s$ axis with zero $x_s$ component of velocity. The sub-satellite point, having coordinates $(0, 0, H - r_e)$, has velocity components $(0, \omega_1 r_e, \dot{H})$. Now, substitution of the value for $\psi$ (into Equation (8) while preserving the signs of the numerator and denominator) results in $\omega_1 = -\sqrt{\left(\dot{\hat{\phi}}\right)^2 + \left(\dot{\hat{\theta}}\sin\hat{\phi}\right)^2}$ showing that the direction of motion of points is towards the $-y_s$ axis, in keeping with the requirement that the $+y_s$ axis points in the direction of the real track.

**Image Memory to Satellite Coordinate System**

Given coordinates $(\tau_m, t_k)$ in image memory for a point target with zero Doppler speed, that target's coordinates $(x_{s0}, y_{s0}, z_{s0})$ in the satellite coordinate system are obtained by requiring that they be located at the intersection of three surfaces [7]:

$$v_d = -(m_x x_{s0} + m_y y_{s0} + m_z z_{s0})/(c\tau_m/2) \quad (11)$$

which is the Doppler surface (with $v_d = 0$ for zero Doppler); with

$$x_{s0}^2 + y_{s0}^2 + z_{s0}^2 = (c\tau_m/2)^2 \quad (12)$$

being the surface of a spherical wave of slant range radius $(c\tau_m/2)$; and

$$x_{s0}^2 + y_{s0}^2 + (z_{s0} - H)^2 = r_e^2 \quad (13)$$

being the surface of an earth sphere radius $r_e$, with orbit radius $H$ on the $z_s$ axis. This intersection point in the satellite coordinate system is

$$\underline{X}_{s0} = \begin{bmatrix} x_{s0} \\ y_{s0} \\ z_{s0} \end{bmatrix} = \begin{bmatrix} \pm\sqrt{(c\tau_m/2)^2 - z_{s0}^2 - y_{s0}^2} \\ -(\dot{H}/(H\omega_1))z_{s0} \\ ((c\tau_m/2)^2 + H^2 - r_e^2)/2H \end{bmatrix} \quad (14)$$

For the case of a circular orbit segment $\dot{H}$ is zero, so that

$$\begin{bmatrix} x_{s0} \\ y_{s0} \\ z_{s0} \end{bmatrix} = \begin{bmatrix} \pm\sqrt{(c\tau_m/2)^2 - z_{s0}^2} \\ 0 \\ ((c\tau_m/2)^2 + H^2 - r_e^2)/2H \end{bmatrix} \quad (15)$$

In both equations $x_{s0}$ is positive for a portside radar (e.g. radar clock angle of 270°), and negative for a starboard radar (e.g. Seasat clock angle of 90°)(see **Satellite Coordinate System** above).

For a circular $(\dot{H} = 0)$ orbit, $y_{s0} = -(H/(\dot{H}\omega_1))z_{s0} = 0$, so that all zero Doppler points lie along the $x_{s0}$ axis for this case. The effect of a non-spherical orbit is therefore grasped by the appearance of a non-zero $y_{s0}$ coordinate in the satellite coordinate system.

**Image Memory to Earth Coordinate System**

The coordinates of an image memory location $(\tau_m, t_k)$ in the earth coordinate system are obtained from Equation (4) to give $\underline{X}_e(\tau_m, t_k)$
In component form this is:

$$\begin{bmatrix} x_e \\ y_e \\ z_e \end{bmatrix} = \begin{bmatrix} x_{s0}(\cos\hat{\phi}\cos\hat{\theta}\cos\psi + \sin\hat{\theta}\sin\psi) + y_{s0}(-\cos\hat{\phi}\cos\hat{\theta}\sin\psi + \sin\hat{\theta}\cos\psi) + (H - z_{s0})\sin\hat{\phi}\cos\hat{\theta} \\ x_{s0}(\cos\hat{\phi}\sin\hat{\theta}\cos\psi - \cos\hat{\theta}\sin\psi) + y_{s0}(-\cos\hat{\phi}\sin\hat{\theta}\sin\psi - \cos\hat{\theta}\cos\psi) + (H - z_{s0})\sin\hat{\phi}\sin\hat{\theta} \\ -x_{s0}\sin\hat{\phi}\cos\hat{\theta} + y_{s0}\sin\hat{\phi}\sin\psi + (H - z_{s0})\cos\hat{\phi} \end{bmatrix} \quad (16)$$

Having obtained $(x_e, y_e, z_e)$, the latitude and longitude associated with this vector can be determined (see **Earth Coordinate System**). The values of $(x_{s0}, y_{s0}, z_{s0})$ are given in Equations (14) or (15) for a particular $(\tau_m, t_k)$, depending on whether the orbit has a significant vertical velocity $(\dot{H} \neq 0)$ or is piecewise spherical $(\dot{H} = 0)$. A two dimensional resampling of a resulting image (now with latitude-longitude coordinates) is required to convert it to a specific map projection.
To resample image memory to produce an image on a final map projection directly, the inverse transformation is needed that locates the image memory coordinates $(\tau_m, t_k)$ for a particular $\underline{X}_e$. The essential difficulty is that there is no straightforward way of determining the orbital time $t_k$ at which a given target at $(\theta_e, \phi_e)$ has zero Doppler, and is not addressed in this paper.

**Radar Coordinate System to Satellite Coordinate System**

This transformation is required to determine the amount of radar illumination for signal processing the SAR data [6]. Any vector in the radar coordinate system (and by association the beam power of that vector) must be transformed into the satellite coordinate system. There are two uses for this transformation.
The first is to map radar beam power levels (the location of the -3dB and 0dB beam power levels in the case of Seasat processing) to signal memory coordinates $(\tau, \eta)$ by transforming the radar coordinates of these power levels first to the satellite coordinate system followed by the transformation to signal memory. The details of the method are shown in **Beam Lines** below.
Secondly, the inverse transformation into the radar coordinate system from the satellite coordinate system can be used to determine the beam illumination at any section along

417the trajectory in signal memory of a particular point target. For each vector in the radar coordinate system it is possible to assign a unique energy flux crossing the unit sphere. This capability is needed to allow for beam illumination effects in the signal processing of the SAR data [4].

The transformation of the vector

$$\underline{X}_a = \begin{bmatrix} x_a \\ y_a \\ z_a \end{bmatrix} \tag{17}$$

into the satellite system coordinate vector $\underline{X}_{sa}$ requires the satellite attitude angles $(\theta_r, \theta_p, \theta_y)$ for roll, pitch and yaw; the deviation angle $H_e$ of the spacecraft nominal track velocity vector from the ground track heading ($H_e = 0$ for non-rotating earth); and the antenna boresight elevation angle $\gamma$ ($-20°$ for Seasat) and clock angle $\xi$ (90° for Seasat).

The transformed coordinates $\underline{X}_{sa} = \underline{N}^T \underline{X}_a$ are obtained by sequential rotations of the satellite coordinate system [8] as follows:

(a) A rotation of $-H_e + \frac{\pi}{2}$ about the $z_s$ axis to align the resulting $x$ axis to the nominal spacecraft heading ($-H_e$), and the resulting $y$ axis starboard ($+\frac{\pi}{2}$) of the satellite;

(b) Sequential rotations $\theta_r, \theta_p, \theta_y$ about the roll ($x$), pitch ($y$) and yaw ($z$) axes to arrive at an attitude independent coordinate system. For Seasat these angles were expected to be in the range of $\pm 1°$ so that $\sin\theta \cong \theta$, $\cos\theta \cong 1$ resulting in a maximum fractional error of $5 \times 10^{-5}$. Note that the yaw angle is in the plane that also defines the deviation angle $H_e$ and so that $H_e$ and yaw measurements are directly coupled;

(c) A rotation of $\xi - \frac{\pi}{2}$ about the $z$ axis, followed by a rotation of $\gamma$ about the $x$ axis to align the resulting $z$ axis with the antenna boresight vector. Note that for a clock angle $\xi = 90°$ (as in Seasat-A) the first rotation about the $z$ axis is zero, and that $\gamma$ is the inclination of the boresight from the vertical ($-20°$ for Seasat).

The inverse transformation is $\underline{X}_a = \underline{N}\, \underline{X}_{sa}$ with $\underline{N}$ given by (including an arbitrary clock angle $\xi$)

$$\underline{N} = \begin{bmatrix}
\sin\left(\frac{\pi}{2} - \xi + H_e\right) & \cos\left(\frac{\pi}{2} - \xi + H_e\right) & -\theta_p \sin\xi - \theta_r \cos\xi \\
-\theta_y \cos\left(\frac{\pi}{2} - \xi + H_e\right) & +\theta_y \sin\left(\frac{\pi}{2} - \xi + H_e\right) & \\
& & \\
-\cos\left(\frac{\pi}{2} - \xi + H_e\right)\cos\gamma & \sin\left(\frac{\pi}{2} - \xi + H_e\right)\cos\gamma & \sin\gamma \\
-\theta_y \sin\left(\frac{\pi}{2} - \xi + H_e\right)\cos\gamma & -\theta_y \cos\left(\frac{\pi}{2} - \xi + H_e\right)\cos\gamma & -(\theta_p \cos\xi - \theta_r \sin\xi)\cos\gamma \\
+(\theta_p \sin H_e + \theta_r \cos H_e)\sin\gamma & +(\theta_p \cos H_e - \theta_r \sin H_e)\sin\gamma & \\
& & \\
\cos\left(\frac{\pi}{2} - \xi + H_e\right)\sin\gamma & -\sin\left(\frac{\pi}{2} - \xi + H_e\right)\sin\gamma & \\
+\theta_y \sin\left(\frac{\pi}{2} - \xi + H_e\right)\sin\gamma & +\theta_y \cos\left(\frac{\pi}{2} - \xi + H_e\right)\sin\gamma & \cos\gamma \\
+(\theta_p \sin H_e + \theta_r \cos H_e)\cos\gamma & +(\theta_p \cos H_e - \theta_r \cos H_e)\cos\gamma & +(\theta_p \cos\xi - \theta_r \sin\xi)\sin\gamma
\end{bmatrix} \tag{18}$$

## Interface with SAR Processor

For signal processing the SAR radar return data, the equation for a target that has zero Doppler at $(\tau_m, t_k)$ must be known as a function of slow time $\eta$ relative to $t_k$ in signal memory. In addition, the Doppler frequency along a signal memory point trajectory is required. The extent of illumination of these signal memory trajectories (in the case of Seasat the loci of the 0dB and -3dB levels) is also required. Depending on the accuracy of the spacecraft attitude data, it may be necessary to obtain estimates of the roll, pitch and yaw angles separately from Doppler centroid frequency as a function of slant range. In the following subsections expressions are derived for each of these functions.
The detailed requirements for these geometrical data depend on the SAR processor design, and here were developed for the first digital Seasat SAR processor [1][3][4].

## Slant Range (Fast Time) Versus Slow Time

Working with the slant range squared, at time $t$ this is given by:

$$\underline{X}_s^T \underline{X}_s = \underline{X}_{s0}^T \underline{X}_{s0} + 2\underline{X}_{s0}^T \left[\underline{\underline{M}}(t_k)\{\underline{B}(t_k) - \underline{B}(t)\}\right] + \left[\underline{B}(t_k) - \underline{B}(t)\right]^T \left[\underline{B}(t_k) - \underline{B}(t)\right] \quad (19)$$

where $t_k$ is the slow time at which the target has zero Doppler frequency at slant range $\sqrt{\underline{X}_{s0}^T \underline{X}_{s0}}$. The earth coordinates for this target are given by (see Equation (4))

$$\underline{X}_e(\tau_m, t_k) = \underline{\underline{M}}^T(t_k)\underline{X}_{s0}(\tau_m) + \underline{B}(t_k) \quad (20)$$

Substitution into Equation (1) describes the trajectory of this earth-fixed point, and using $\underline{X}^T \underline{X} = \underline{X}\, \underline{X}^T$ gives Equation (19). This expression can be reduced to differential form, by substituting $t = t_k + \eta$ and Taylor expanding $\underline{B}(t_k + \eta)$ to give

$$\underline{X}_s^T \underline{X}_s = \underline{X}_{s0}^T \underline{X}_{s0} + \left[\underline{\dot{B}}^T(t_k)\underline{\dot{B}}(t_k) - \underline{X}_{s0}^T \underline{\underline{M}}(t_k)\underline{\ddot{B}}(t_k)\right]\eta^2 + \cdots \quad (21)$$

Since $\underline{X}_s^T \underline{X}_s = (c\tau/2)^2$ and $\underline{X}_{s0}^T \underline{X}_{s0} = (c\tau_m/2)^2$ are the slant ranges squared at slow time $t = t_k + \eta$ and at zero Doppler slow time $t_k$ respectively, this form is the conventional hyperbolic locus [7]. Higher order terms in $\eta$ may be required if more accurate expressions for slant range are found to be necessary. Regarding Seasat, an orbital analysis of the fast and slow time invariance showed the coefficient of $\eta^2$ to be sufficiently slowly varying that this hyperbolic expansion is justified, and appropriate for block-wise azimuth compression [4].

The above expression for slant range does not involve a term linear in $\eta$. Because $t_k$ is the slow time at which the target has a zero Doppler, the term $\left[\underline{\underline{M}}(t_k)\underline{\dot{B}}(t_k)\right]^T \underline{X}_{s0}\eta$ appearing in the calculation of the differential form is precisely equal to zero (see Equation (6) where $v_d$ is defined in terms of $\left(\underline{\underline{M}}\, \underline{\dot{B}}\right)^T \underline{X}_s$; at zero Doppler $v_d = 0$, and the result follows). Putting

$$\Omega(\tau_m, t_k) = \underline{\dot{B}}^T(t_k)\underline{\dot{B}}(t_k) - \underline{X}_{s0}^T(\tau_m, t_k)\underline{\underline{M}}(t_k)\underline{\ddot{B}}(t_k) \tag{22}$$

the parabolic approximation is given by

$$\frac{c\tau}{2} = \frac{c\tau_m}{2} + \frac{\Omega}{c\tau_m}\eta^2 \tag{23}$$

using the leading term in the binominal expansion of

$$c\tau/2 = c\tau_m/2\ \sqrt{1 + \Omega(2\eta/c\tau_m)^2}. \tag{24}$$

Expanding $\Omega(\tau_m, t_k)$ we obtain

$$\begin{aligned}
\Omega/H^2 &= \dot{H}^2/H^2 + \dot{\hat{\theta}}^2 \sin^2\hat{\phi} + \dot{\hat{\phi}}^2 \\
&+ \dot{\hat{\theta}}^2\sin^2\hat{\phi}\left[\cot\hat{\phi}\left(\frac{x_{s0}}{H}\cos\psi - \frac{y_{s0}}{H}\sin\psi\right) - \frac{z_{s0}}{H}\right] - \dot{\hat{\phi}}^2\frac{z_{s0}}{H} + 2\dot{\hat{\theta}}\,\dot{\hat{\phi}}\cos\hat{\phi}\left(\frac{x_{s0}}{H}\sin\psi + \frac{y_{s0}}{H}\cos\psi\right) \\
&+ \ddot{\hat{\theta}}\sin\hat{\phi}\left(\frac{x_{s0}}{H}\sin\psi + \frac{y_{s0}}{H}\cos\psi\right) - \ddot{\hat{\phi}}\left(\frac{x_{s0}}{H}\cos\psi - \frac{y_{s0}}{H}\sin\psi\right) \\
&+ 2\frac{\dot{H}}{H}\left[\dot{\hat{\theta}}\sin\hat{\phi}\left(\frac{x_{s0}}{H}\sin\psi + \frac{y_{s0}}{H}\cos\psi\right) - \dot{\hat{\phi}}\left(\frac{x_{s0}}{H}\cos\psi - \frac{y_{s0}}{H}\sin\psi\right)\right] + \frac{\ddot{H}}{H}\frac{z_{s0}}{H}
\end{aligned} \tag{25}$$

A piecewise circular orbit requires (see Equations (14), (15)), $\dot{H} = 0, y_{s0} = 0$. In general numerical analysis is required to determine if any acceleration terms are needed for the orbit used. For Seasat no acceleration terms were required.

**Isodoppler Lines**

In fast time slow time coordinates $(\tau, \eta)$, with $\eta = 0$ corresponding to orbital time $t_k$ (slow time for zero Doppler), the locus of points with a given Doppler $v_d$ in signal memory is found from

$$\tau = \tau_m\left(1 + v_d^2/2\Omega(\tau_m, t_k)\right) \tag{26}$$

and

$$\eta = (c\tau_m/2)v_d/\Omega(\tau_m, t_k) \tag{27}$$

where $\tau_m$ is the fast time (slant range = $c\tau_m/2$) at zero Doppler ($v_d = 0$) and $\Omega$ is given in Equation (25).

These equations are parametric in $\tau_m$, variations in $\tau_m$ generating a family of coordinates $(\tau, \eta)$ for a fixed $v_d$, and hence an isodoppler line. The expression relating Doppler to slow time is obtained by noting that the Doppler speed is the rate of change of slant range with respect to variations in slow time, i.e. $v_d = d(c\tau/2)/d\eta$. Performing the differentiation on the parabolic expansion for slant range (Equation (23)) results in Equation (27). Substituting this value for $\eta$ into the Equation (23) gives Equation (26). In this formulation, approaching velocities, leading to decrements in range, are negative. Then the corresponding equations in terms of Doppler frequencies [4] $f_d = -2f_0 v_d/c$ are

$$\tau = \tau_m\left(1 + c^2 f_d^2/(8f_0^2\Omega)\right) \tag{28}$$

and

$$\eta = -(c\tau_m/2)(c/(2f_0\Omega))f_d \tag{29}$$

where $f_0$ is the radar frequency.

Since there is a one-to-one correspondence between slow time $\eta$ and $f_d$, the equation relating fast time to Doppler frequency identifies trajectories of targets in the slant range - Doppler frequency space.

## Beam Lines

Each target following the fast-time slow-time trajectory in signal memory is illuminated by the radar beam according to the orientation of the beam with respect to the real track. For instance, if the radar beam happens to have its boresight vector in a plane perpendicular to the real track, the point of zero Doppler will be illuminated. In general, due to the combined effects of earth rotation and spacecraft attitude variation, sections of trajectory not encompassing zero doppler may very well be illuminated.

In fast-time slow-time coordinates (with $\eta = 0$ corresponding to an orbital time $t_k$), the signal memory coordinates $(\tau, \eta)$ at which a point target, that has zero doppler at $(\tau_m, 0)$ is illuminated by a radar beam vector $\underline{X}_a$ is given by

$$\frac{c\tau}{2} = \frac{x_{s0}\left[\left(\frac{z_{s0}}{H}\right)-1\right]+z_{s0}^2\frac{\omega_3}{H\omega_1}}{x_{sa}\left[\left(\frac{z_{s0}}{H}\right)-1\right]+z_{sa}z_{s0}\frac{\omega_3}{H\omega_1}} \tag{30}$$

and

$$\eta = \frac{\left(\frac{c\tau}{2}\right)y_{sa}+\frac{\dot{H}z_{s0}}{H\omega_1}}{x_{s0}\omega_3+\omega_1(H-z_{s0})} \tag{31}$$

Here $x_{so} = r_{s0}\cos\theta$, $z_{s0} = r_{s0}\sin\theta$ are the satellite system coordinates of the target at zero Doppler where

$$\tan\theta = \frac{z_{sa}+2\frac{\dot{H}}{c\tau}\left(\frac{z_{s0}}{H}-1\right)\eta}{x_{sa}-2\frac{\omega_3 \dot{H} z_{s0}}{\omega_1 H\,c\tau}\eta} \tag{32}$$

and

$$r_{s0} = Hk(\theta)\sin\theta \pm \sqrt{r_e^2 - (H\cos\theta)^2 - \left(\dot{H}/\omega_1\right)^2(\sin\theta\tan\theta)^2 - \left(\dot{H}/\omega_1\right)^2(\sin\theta\tan\theta)^2} \tag{33}$$

The expressions for $\omega_1, \omega_3$ are defined in Equation (8) and

$$k(\theta) = \sqrt{1 - \left(\frac{\dot{H}}{H\omega_1}\tan\theta\right)^2} \tag{34}$$

The components $(x_{sa}, y_{sa}, z_{sa})$ of the vector $\underline{X}_{sa}$ are calculated using the transpose of the matrix $\underline{N}$ (Equation (18)) operating on $\underline{X}_a$. For a starboard antenna (Seasat-A clock angle $\xi = 90°$) choose the negative square root in Equation (33); otherwise for a portside antenna choose the positive root for ($\xi = 270°$).

These equations for $(\tau, \eta)$ do not have a simple closed form solution for $\dot{H} \neq 0$, due to the coupling of terms in $\dot{H}, \tau, z_{so}$ and $\eta$ in the numerator and denominator of $\tan\theta$. For $\dot{H} = 0$, the values for $\tan\theta$ (Equation (32), $r_{s0}$ (Equation (33)), $\tau$ (Equation (30)) and $\eta$ (Equation (31) can be immediately found. For $\dot{H} \neq 0$ an iterative method must be used, based on calculating a sequence of values $\theta, r_{s0}, \tau, \eta$ beginning with a piecewise circular orbit fit $(\dot{H} = 0)$ to the elliptical orbit.

These equations for $(\tau, \eta)$ are essentially parametric in $\underline{X}_a$, so that each radar beam vector is assigned a unique point in signal memory. For Seasat, with the radar beam following the 0, -3dB beam profiles (along track angle $\beta$ equal to 0°, ±0.52°, across track angle $\alpha$ variable) the times at which the point target crosses these beam profiles can be plotted. In the satellite coordinate system, the region illuminated by the beam does not change position unless attitude, clock or elevation angles vary, or a sufficiently different section of the sub-satellite track is involved. The degree to which this region can be assumed invariant of slow time should be taken into account depending on the radar beam geometry, and the time variation of $H_e$.

Equations (30) and (31) result from requiring that, at time $(t_k + \eta)$, the radar beam vector unit vector $\underline{X}_{sa}$ must have precisely the same direction cosines as the target with zero Doppler coordinates $\underline{X}_{s0}$ at the time $t_k$. To first order in $\eta$ the vector equivalence is

$$(c\tau/2)\underline{X}_{sa} = \underline{X}_{s0} + \underline{\dot{X}}_{s0}\eta \tag{35}$$

with $\underline{\dot{X}}_{s0}$ given by Equation (9), and must be evaluated at slow time $t_k$ for a slant range at zero Doppler of $c\tau_m/2$. Likewise the vector $\underline{X}_{s0}$ is obtained from the transformation of zero Doppler image memory coordinates $(\tau_m, t_k)$ to satellite coordinates (Equation (15)). It is then straightforward to compute the radar beam vector $\underline{X}_{sa}$ in the satellite system that corresponds to any $(\tau, \eta)$. A further transformation $\underline{X}_a = \underline{\underline{N}}\,\underline{X}_{sa}$ (with $\underline{\underline{N}}$ as in Equation (18)) determines the radar beam vector in the radar coordinate system taking into account attitude angles. This procedure then allows immediate calculation of the beam illumination at any point $(\tau, \eta)$ along the target loci in signal memory.

**Attitude Measurements from Range Doppler Values**

The method of intersecting surfaces can also be used (Equations (11), (12), (13)) to obtain the satellite system coordinates of a point with slant range $c\tau/2$ at any Doppler speed $v_d$. The satellite system unit vector for such a point is

$$\begin{bmatrix} \hat{x}_v \\ \hat{y}_v \\ \hat{z}_v \end{bmatrix} = \begin{bmatrix} \pm\sqrt{1 - \hat{y}_v^2 - \hat{z}_v^2} \\ -(v_d + \dot{H}\hat{z}_v)/(H\omega_1) \\ \frac{H}{c\tau} + \frac{c\tau}{4H} - \frac{r_e^2}{Hc\tau} \end{bmatrix} \tag{38}$$

It has also been shown that the transformation $\underline{X}_a = \underline{\underline{N}}\,\underline{X}_{sa}$ (Equation (18)) gives the components of this point in the radar coordinate system, provided the attitude angles that define $\underline{\underline{N}}$ are known.

If the attitude angles are not known to sufficient precision, a reverse procedure can be used to improve that precision, because it is possible to find a set of attitude dependent radar beam vectors that correspond to a set of Doppler centroid frequencies for different slant ranges $c\tau/2$. In this case azimuth frequency analysis of the range compressed data is used to determine the Doppler centroid frequency $f_d = -2f_0 v_d/c$ for a set of slant ranges $c\tau/2$ and this is input to the algorithm described in what follows.

The attitude angles are assumed to vary slowly compared with the PRF, so that azimuth processing of the slant range signal for a set of slant ranges $c\tau/2$ will achieve a useful precision for the Doppler centroid frequency. The actual determination of Doppler centroid frequencies [4] is not addressed in this paper.

The components of a unit vector in the radar coordinate system transformed to the satellite coordinate system are

$$\begin{bmatrix} x_{sa} \\ y_{sa} \\ z_{sa} \end{bmatrix} = \begin{bmatrix} -y_a \cos\gamma \cos\left(\frac{\pi}{2} - \xi + H_e\right) + z_a \sin\gamma \cos\left(\frac{\pi}{2} - \xi + H_e\right) \\ +\theta_y\left\{-y_a \cos\gamma \sin\left(\frac{\pi}{2} - \xi + H_e\right) + z_a \sin\gamma \sin\left(\frac{\pi}{2} - \xi + H_e\right)\right\} \\ +\theta_p\{y_a \sin\gamma \sin H_e + z_a \cos\gamma \sin H_e\} \\ +\theta_r\{y_a \sin\gamma \cos H_e + z_a \cos\gamma \cos H_e\} \\ \\ y_a \cos\gamma \sin\left(\frac{\pi}{2} - \xi + H_e\right) - z_a \sin\gamma \sin\left(\frac{\pi}{2} - \xi + H_e\right) \\ +\theta_y\left\{-y_a \cos\gamma \sin\left(\frac{\pi}{2} - \xi + H_e\right) + z_a \sin\gamma \cos\left(\frac{\pi}{2} - \xi + H_e\right)\right\} \\ +\theta_p\{y_a \sin\gamma \cos H_e + z_a \cos\gamma \cos H_e\} \\ +\theta_r\{-y_a \sin\gamma \sin H_e - z_a \cos\gamma \sin H_e\} \\ \\ y_a \sin\gamma + z_a \cos\gamma \\ +\theta_p\{-y_a \cos\gamma \cos\xi + z_a \sin\gamma \cos\xi\} \\ +\theta_r\{y_a \cos\gamma \sin\xi - z_a \sin\gamma \sin\xi\} \end{bmatrix} \quad (39)$$

where $y_a = \sin\alpha$ and $z_a = \cos\alpha$ are the unit vector components of the radar beam (Figure 5) perpendicular to the along track direction. Note that $y_a^2 + z_a^2 = 1$, and that $x_a = 0$ forces the radar vector to be along the 0dB line in the case of Seasat.

We now equate the $\hat{z}_v$ component of the unit vector in Equation (38) to the $z_{sa}$ component of the unit vector in Equation (39) to give

$$\hat{z}_v = \cos(\alpha - \gamma) + (\theta_p \cos\xi + \theta_r \sin\xi)\sin(\alpha - \gamma) \quad (40)$$

This equation is used to solve for $\alpha$ for small attitude angles so that terms $\sim (\theta_p \cos\xi + \theta_r \sin\xi)^2$ can be neglected, giving

$$\alpha = \sin^{-1}\hat{z}_v + \theta_p \cos\xi + \theta_r \sin\xi + \gamma - \frac{\pi}{2} \quad (41)$$

In this development only the z component of the radar beam defines the across track angle $\alpha$, and it is the only component that is explicitly dependent on fast time $\tau$ through $\hat{z}_v$ in Equation (38). As an illustration, consider $\theta_p = \theta_r = \theta_y = 0$, and with the radar beam pointing directly at the earth's centre ($\gamma = 0$). This gives $\hat{z}_v = 1$ and $\sin^{-1}\hat{z}_v = \frac{\pi}{2}$. For this case the across track angle $\alpha = 0$ for Equation (41) as expected.

The component $\hat{y}_v$ contains the dependence on $v_d$, so equating the two $y$ components of Equations (38) and (39) the Doppler speed is

$$v_d = -\dot{H}\hat{z}_v - H\omega_1\{\sin\left(\frac{\pi}{2} - \xi + H_e\right)\sin(\alpha - \gamma) + \theta_y \cos\left(\frac{\pi}{2} - \xi + H_e\right)\sin(\alpha - \gamma) -$$

$$\theta_p \cos H_e \cos(\alpha - \gamma) - \theta_r \sin H_e \cos(\alpha - \gamma)\} \tag{42}$$

In order to apply these equations to the case of Seasat, the clock angle $\xi = 90°$, so that

$$\alpha = \sin^{-1}\hat{z}_v(\tau) + \theta_r + \gamma - \frac{\pi}{2} \tag{43}$$

and

$$v_d = -\dot{H}\hat{z}_v - H\omega_1\{\sin H_e \sin(\alpha - \gamma) + \theta_y \cos H_e \sin(\alpha - \gamma) - \theta_p \cos H_e \cos(\alpha - \gamma) -$$

$$\theta_r \sin H_e \cos(\alpha - \gamma)\} \tag{44}$$

This equation is linear in $\theta_y$ and $\theta_p$ but non-linear in $\theta_r$.

As mentioned above, the attitude angles are expected to be constant over the time range of the Doppler centroid values. During the slow time range of the set of Doppler centroid values the orbital parameters $\dot{H}, H, \omega_1$ and $H_e$ are to be fixed at the midpoint of the time range, that is, at a slow time $t_k$. For Seasat, assuming 512 azimuth samples for a typical PRF of 1540 Hz, this time range is about 1/3 second. To obtain the Doppler centroid values, the range-compressed data is segmented into separate slant range groups each centered at a particular $c\tau/2$. This results in a set of Doppler centroid frequencies $f_d \pm \delta f_d$ and associated slant ranges $c\tau/2$. Given a sufficient number of samples a single bright target starting at a large slant range will have a Doppler centroid frequency that decreases as the bright target approaches its point of closest approach. It is this trajectory that Equation (44) defines and the next step is to describe the attitude angle estimation process.

In general Equation (44) is non-linear in the attitude parameters and a suitable technique for solving for these parameters is an iterative Maximum Likelihood Method (MLM)[11]. The starting values of the attitude parameters can be made as follows, to remove the non-linearity in $\theta_r$, and allow for a fast initial Least Squares estimate. Making the small angle approximation for $\theta_r$ in both $\cos(\alpha - \gamma)$ and $\sin(\alpha - \gamma)$ gives

$$\cos(\alpha - \gamma) = \theta_r\sqrt{1 - \hat{z}_v^2} + \hat{z}_v \tag{45}$$

$$\sin(\alpha - \gamma) = \theta_r\hat{z}_v - \sqrt{1 - \hat{z}_v^2}\,\hat{z}_v \tag{46}$$

so that Equation (42) becomes linear to first order for small yaw, pitch and roll angles:

$$v_d = -\dot{H}\hat{z}_v + H\omega_1\{\sqrt{1 - \hat{z}_v^2}\sin H_e - \theta_y\sqrt{1 - \hat{z}_v^2}\cos H_e - \theta_p\hat{z}_v \cos H_e - 2\theta_r\hat{z}_v \sin H_e\} \tag{47}$$

This expression for the Doppler centroid as a function of slant range (through $\hat{z}_v$ in Equation (38)) is linear in yaw, pitch and roll angles and includes the $\dot{H}$ term allowing for a non-circular orbit. A further simplification for the Least Squares fit is to assume $\dot{H} = 0$; that is, a circular orbit.

Using the initial values obtained from the Least Squares fit, the MLM will result in a set of attitude angles and their errors, using Equation (44) as fitting function.

**Deviation Angle $H_e$**

In order to modify the LANDSAT 1 orbital model [6] to the small eccentricity of the Seasat orbit, simple expressions for the true anomaly $\rho$ and orbit height $H$ from the earth centre were derived and were used in testing before the SAR processor was operational. Assuming the orbit plane angular frequency $\omega_s$ of the satellite about the point midway between the foci to be constant (due to the small difference between the major and minor semi-axes referred to above in **Inertial Coordinate System**)

$$\tan \rho = \sqrt{1 - e^2} \sin(\omega_s t + \chi) / \cos(\omega_s t + \chi) - e \qquad (48)$$

and

$$H = a[1 - e \cos(\omega_s t + \chi)] \qquad (49)$$

where $e$ is the eccentricity, $a$ is the semi-major axis and $\chi$ is a phase angle to control the position of orbit perigee. In terms of this model, with orbit inclination $\varepsilon$ relative to the north polar axis, the satellite's spherical latitude coordinate angle $\hat{\phi}$ (see Equation (2)) is given by

$$\cos \hat{\phi} = \cos \rho \cos \varepsilon \qquad (50)$$

For Seasat $\omega_s = 0.059553°/\text{second}$, and $\varepsilon = 18°$ and the deviation of the real track from the nominal heading is given by

$$\tan H_e = \{(\omega_s/\omega_e) \cos \varepsilon \sin \rho\}/\{1 + (\omega_s/\omega_e) \sin \varepsilon\} \qquad (51)$$

Latitude is unaffected by Earth rotation, so the phase angle can be determined from the satellite's ECEF $\hat{\phi}$ to give

$$\tan H_e = \{(\omega_s/\omega_e)\sqrt{\cos^2\varepsilon - \cos^2\hat{\phi}}\}/\{1 + (\omega_s/\omega_e) \sin \varepsilon\} \qquad (52)$$

It is also possible to use the satellite's ECEF coordinates to determine the angle $H_e$. Treating the ECEF system as fixed, the inertial system rotates from east to west around the polar ($z_e$) axis with a spin vector $-\underline{\omega}_e$. The satellite velocity vectors in the two coordinate systems are then related by the equation [9]

$$\underline{\dot{B}}_{IN}(t) = \underline{\dot{B}}(t) + \underline{\omega}_e \times \underline{B}(t) \qquad (53)$$

where $\underline{B}(t)$ and $\underline{\dot{B}}(t)$ are the spacecraft's ECEF position and velocity vectors (Equation (2) and the first time derivative); and $\underline{\dot{B}}_{IN}(t)$ is the inertial system's velocity vector of the spacecraft. The heading deviation is the angle between these two vectors, where $H_e$ is the angle taken from $\underline{\dot{B}}(t)$ to $\underline{\dot{B}}_{IN}(t)$ and is given by [8]

$$\cos H_e = \{\underline{\dot{B}}(t) \cdot \underline{\dot{B}}_{IN}(t)\} / \sqrt{\left|\underline{\dot{B}}(t)\right|^2 + \left|\underline{\dot{B}}_{IN}(t)\right|^2} \qquad (54)$$

This alternative method was not used for the production of the first digital (Trois Rivieres) Seasat SAR image [1].

**SUMMARY AND CONCLUSIONS**

This paper is a detailed and historical review of the development of the geometric model used in the production the first digitally processed Seasat-A SAR image. The concepts and equations were incorporated into the early MacDonald Dettwiler [4] digital processors (e.g. GSAR) and provided the azimuth compression frequency, the geometry to take into account the range-Doppler track curvature and illumination, as well as a technique for determining the spacecraft attitude angles. Because the image memory coordinate system is pivotal in the technique, look registration is automatically taken into account, and the transformation of the image to any map projection integral to the procedure. The image location accuracy was subsequently assessed using the UTM coordinates of targets in several scenes resulting in a standard deviation (RMS) of approximately 30 metres [10]. The essence of the geometric model described here has played a significant role in the subsequent development of the Radarsat ground processors developed for the Canadian government.

**ACKNOWLEDGEMENTS**

The author wishes to acknowledge Keith Raney and John Bennett for introducing the problem of spaceborne SAR processing, and that first MDA SAR team [3] for their constructive dialogue, especially the late Robert Deane. The late Roy Selby was indispensible in searching for errors in the algebra. John Macdonald and the late Dave Sloan were very supportive throughout this work.

26## REFERENCES

(1)  Aviation Week and Space Technology, p19, 26 February 1979.
(2)  SEASAT-A SAR System Design Review, JPL Report, April 1977.
(3)  I.E.E.E. Canadian Review, Fall/Autumn, 2015.
(4)  Cumming, Ian G. and Wong, Frank H; Digital Processing of Synthetic Aperture Radar Data: Algorithms and Implementation; Artech House, Norwood MA. 2005.
(5)  Escobal, Pedro Ramon; Methods of Orbit Determination; Krieger Publishing Company, Malabar FL $2^{nd}$ Ed. 1976.
(6)  Kratky, Vladimir; "Photogrammetric Solution for Precision Processing of ERTS Images"; Presented Paper Commission II, XII International Congress of Photogrammetry, Ottawa, 1972
(7)  Harger, Robert. O.; Synthetic Aperture Systems – Theory and Design; Academic Press, 1970
(8)  Korn, Granino A. and Korn, Theresa M.; Mathematical Handbook for Scientists and Engineers; McGraw-Hill Book Company, New York NY $2^{nd}$ Ed. 1968
(9)  Rutherford, D.E.; Vector Methods Applied to Differential Geometry, Mechanics, and Potential Theory; Interscience Publishers, Inc., New York NY Ninth Ed. 1957.
(10) Seasat SAR Performance Evaluation Study Final Report JPL 00-0676-D00 June 1982: https://cors.archive.org/stream/NASA_NTRS_Archive_19830009652/NASA_NTRS_Archive_19830009652_djvu.txt
(11)  Orear, Jay; Notes on Statistics for Physicists, UCRL-8417, University of California, Radiation Laboratory, Berkeley California; August 13, 1958. See also: Orth, P.H.R.; Falk, W.R. and Jones, G; Use of the maximum likelihood technique, for fitting counting distributions. I. Application to the sum of two exponentials with constant background; Nuclear Instruments and Methods, Volume 65, Issue 3, p. 301-306. (1968)